\documentclass[pra,twocolumn,aps,groupedaddress,longbibliography,nofootinbib]{revtex4-1}
\usepackage{graphicx}
\usepackage{float}
\usepackage{hyperref}
\hypersetup{
      colorlinks=true,
      citecolor=blue,
      linkcolor=blue,
      urlcolor=blue}
\usepackage{amsmath}
\usepackage{amsfonts}
\usepackage{amssymb}
\usepackage{times}

\newcommand{\Ut}{\tilde{U}}
\newcommand{\br}{\mathbf{r}}
\newcommand{\brho}{\boldsymbol{\rho}}
\newcommand{\bk}{\mathbf{k}}
\newcommand{\MF}{\mathcal{F}}
\begin{document}
\title{Infinite dipolar droplet - a simple theory for the macro-droplet regime} 
\author{Sukla Pal, D. Baillie and P. B. Blakie}
\affiliation{Department of Physics, Centre for Quantum Science, and The Dodd-Walls
            	Centre for \\ Photonic and Quantum Technologies, University of
            	Otago, Dunedin 9016, New Zealand}
         \date{\today}
          
\begin{abstract}  	
In this paper we develop a theory for an infinitely long droplet state of a zero temperature dipolar bosonic gas. The infinite droplet theory yields simpler equations to solve for the droplet state and its collective excitations. We explore the behavior of infinite droplets using numerical and variational solutions, and demonstrate that it can provide a quantitative description of large finite droplets  of the type produced in experiments.  
We also consider the axial speed of sound and the  thermodynamic limit of a dipolar droplet.
                  \end{abstract}
         \maketitle

\section{Introduction}
Dipolar bosonic gases of magnetic atoms interact with a long-ranged and anisotropic dipole-dipole interaction (DDI). Experiments using the highly magnetic atoms of dysprosium \cite{Kadau2016a,Ferrier-Barbut2016a,Schmitt2016a} and erbium \cite{Chomaz2016a} have prepared one or several self-bound quantum droplets that cohere, even in the absence of any external confinement. These quantum droplets occur in the dipole-dominated regime, where the short ranged $s$-wave interactions are weaker than the DDIs  \cite{Baillie2016b,Wachtler2016b}, and the overall two-body interactions are attractive. In this regime the effects of quantum fluctuation corrections become important  \cite{LHY1957,Lima2011a,Lima2012a} and stabilize the droplets against mechanical collapse \cite{Petrov2015a,Saito2016a,Wachtler2016a,Bisset2016a}.  
An important effect of the DDIs is that the droplets take a filament shape -- elongating in the direction that the dipoles are polarized -- to minimise the DDI energy.  Due to the long-ranged and anisotropic character of the DDI and the highly elongated droplet shape, quantitative calculations for the droplet state and their excitations are a challenging numerical problem. 

\begin{figure}[H]
  	\centering
  	 \includegraphics[width=0.6\linewidth]{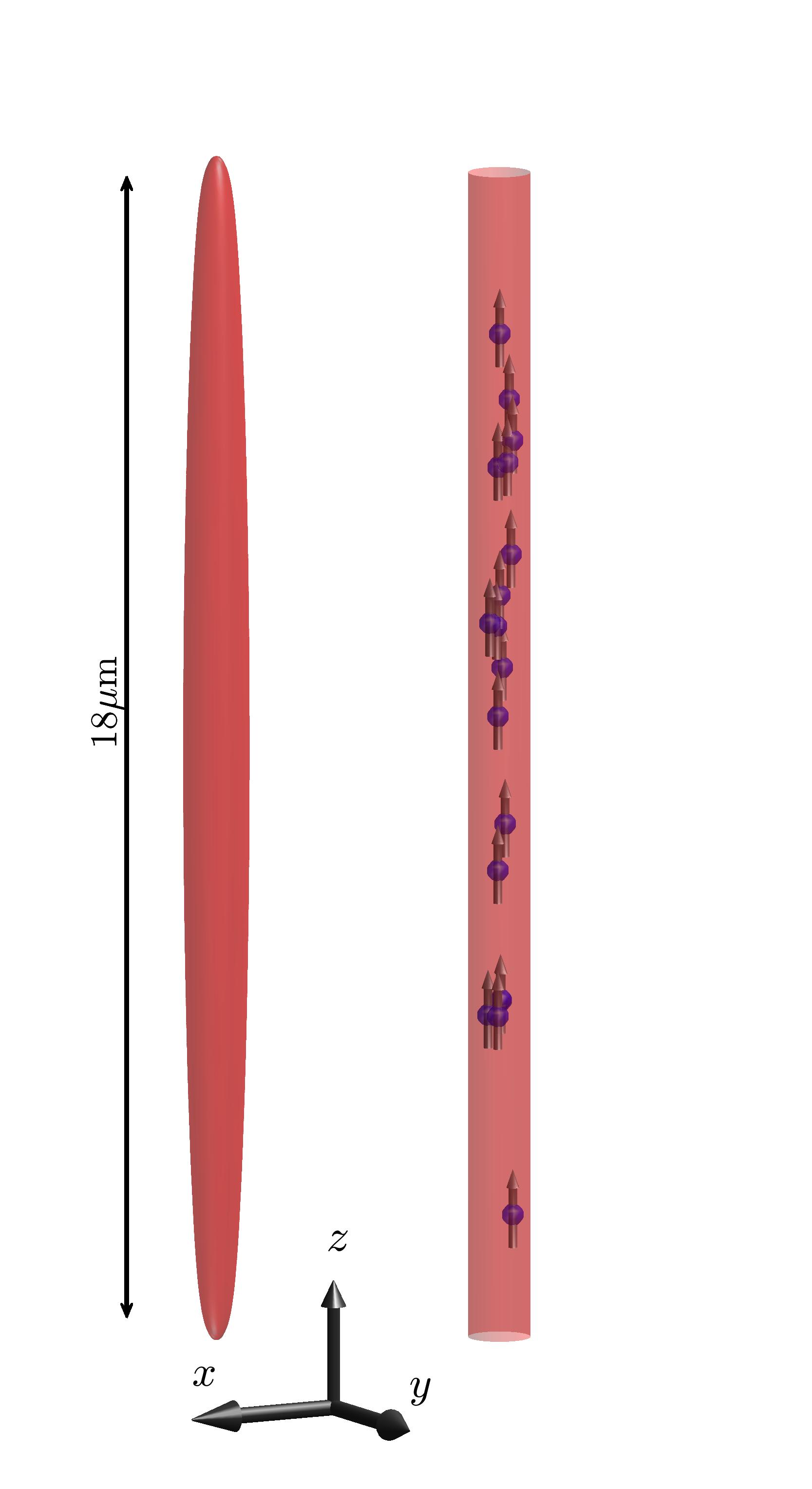}
  	\caption{   (Left) Density isosurface of a finite free-space droplet with $N=2.5\times10^4$  $^{164}$Dy atoms, $a_{dd}=130.8a_0$ and $a_s=80a_0$. (Right) Schematic  of a section of the   infinite dipolar droplet we develop here.}
  	\label{schematic}
  \end{figure}

Here we develop an infinite droplet theory. In this theory we consider the idealized case of a droplet that is an infinitely long filament (along the direction of the dipole orientation) of specified linear density $n$ (see Fig.~\ref{schematic}).  This idealization should be a good approximation to a finite droplet with sufficiently many atoms $N$ that it is highly elongated. Such droplets, often referred to as macro-droplets, have been prepared with $N\sim2\times10^4$ atoms \cite{Chomaz2016a} and can persist as a self-bound droplet even in the absence of any confinement \cite{Schmitt2016a}.

We present results for the infinite droplet using numerical solutions and a variational approximation we develop. We explore the character of the infinite droplet excitations and show that the recently identified ``anti-roton" effect \cite{Pal2020} plays an important role in the spectrum of a droplet.
 We consider a mapping to compare the infinite droplet results to a finite droplet, and use this to make a comparison to the  excitation spectrum of a finite droplet. 
 We find that often the infinite droplet is not mechanically stable, with long wavelength axial modes being dynamically unstable. In these cases the axial speed of sound of the infinite droplet is  imaginary. However, as these instabilities only manifest at wavelengths that are much longer than the length of the finite droplet, they can be considered inconsequential, and the corresponding finite droplet is stable.  Finally, we consider taking a finite droplet to the thermodynamic limit by letting $N\to\infty$, which is different to the infinite dipolar droplet.

\section{Formalism}\label{theory}
\subsection{EGPE  theory}\label{eGPE}
It is now well-established (e.g.~see \cite{Ferrier-Barbut2016a,Schmitt2016a,Chomaz2016a,Baillie2016b,Wachtler2016b,Wachtler2016a,Bisset2016a}\footnote{Alternative theories to the EGPE  have been developed, e.g.~see \cite{Bottcher2019b,Halperin2021a}.}) that the coherent atomic field $\psi(\br)=\langle\hat{\psi}(\br)\rangle$  of a quantum droplet is governed by the extended Gross-Pitaevskii equation (EPGE) that includes the effects of beyond-meanfield quantum fluctuations.
The EPGE takes the form $\mathcal{L} \psi = \mu \psi$, 
where $\mu$ is the chemical potential and
     \begin{align}\label{gp}
     \mathcal{L} &\equiv -\frac{\hbar^2\nabla^2}{2M}+V_\text{tr} +  \int d\br'U(\br-\br')|\psi(\br')|^2 + \gamma_\text{QF}|\psi|^3, \\
    V_\text{tr} &= \frac{1}{2}M\omega_\rho^2(x^2+y^2),
     \end{align}
 are the EGPE  operator and trapping potential, respectively. Here only transversal harmonic confinement is considered with angular frequency $\omega_\rho$, although many of our results are for the free-space case of $\omega_\rho=0$.
 The two-body interactions are described by the interaction potential 
  \begin{align}
  U({\bf r}) = g_s \delta(\br) + \frac{3g_{dd}}{4\pi r^3}\left(1 - \frac{3z^2}{r^2}\right).\label{Urfull}
  \end{align} 
Here $g_s=4\pi a_s\hbar^2/M$ is the short-ranged coupling constant where $a_s$ is the $s$-wave scattering length. The atoms have magnetic dipoles of moment $\mu_m$ aligned along the $z$-axis by an external bias field. The DDIs between the atoms are characterized by the coupling constant $g_{dd} = 4\pi\hbar^2 a_{dd}/M$, where $a_{dd}= \mu_0\mu_m^2 M /12 \pi \hbar^2$.  
The quantum fluctuations are described by the higher order local nonlinear terms with coefficient (see \cite{Lima2011a, Lima2012a, Bisset2016a})
     \begin{eqnarray}
     	\gamma_\text{QF} = \frac{32}{3}g_s\sqrt{\frac{a_s^3}{\pi}} \left( 1+ \frac32 \epsilon_{dd}^2\right),  
     \end{eqnarray}
  where $\epsilon_{dd} = a_{dd}/a_s$. 
  
\subsection{Infinite droplet state}
Here we develop a model for a droplet that is infinitely long and translationally invariant along the $z$-axis, i.e.~
\begin{align}
\psi(\br) =  \sqrt{n} \chi(\rho),\label{decomp}
\end{align}
where $n$ is the specified linear density and $\chi(\rho)$ is the unit normalized transverse mode. Here we have utilized cylindrical symmetry to reduce the dependence of the transverse mode to the radial coordinate $\rho=\sqrt{x^2+y^2}$.
A decomposition similar to Eq.~(\ref{decomp}) was explored previously in Ref.~\cite{Pal2020} (also see \cite{Edmonds2020a}) as a model of a dipolar condensate in an elongated tube trap. Here we extend this theory to include quantum fluctuations and hence the droplet regime which occurs when the DDI interactions dominate over the contact interactions (i.e.~$\epsilon_{dd}>1$). In this regime the droplet can exist (as a transversally localized state) even in the absence of any transverse confinement, i.e.~can become self-bound.
Using (\ref{decomp}), the EGPE  simplifies to 
  \begin{align}
  \mathcal{L} ^\perp\chi=\mu\chi,\label{GPE}
  \end{align}
   where 
    \begin{align}
      \mathcal{L}^{\perp}   \equiv& -\frac{\hbar^2}{2M\rho}\frac{\partial}{\partial \rho} \left(\rho\frac{\partial }{\partial \rho} \right)+ V_\text{tr} +  n(g_s-g_{dd})|\chi(\rho)|^2  \nonumber \\
      &+n^{3/2}\gamma_\text{QF}|\chi({ {\rho}})|^3,\label{LGPperp}
  \end{align} 
   Notice that for the infinite droplet state the two-body interactions appear as an effective contact interaction. We can see this from the $k$-space form of the interaction potential [i.e.~Fourier transform of Eq.~(\ref{Urfull})] 
  \begin{align}
  \tilde{U}(\bk)=g_s+g_{dd}\left(\frac{3 k_z^2}{k^2}-1\right),
  \end{align}
  which approaches the constant value $g_s-g_{dd}$ in the $k_z\to0$ limit appropriate to the infinite droplet state.
  
Because Eq.~(\ref{LGPperp}) does not involve an explicit long range interaction term, it avoids many of the technical issues that normally arise in solving the EGPE with DDIs. Thus, the transverse profile of infinite droplet state can be solved for using standard packages for solving the Gross-Pitaevskii equation adapted to include the higher-order local nonlinearity of the quantum fluctuation term.

\subsection{Excitations of the infinite droplet}
Bogoliubov theory provides a description of the quasiparticle excitations of the droplet \cite{Baillie2017a,Petrov2015a}. These excitations can be obtained by linearizing the time-dependent GPE, $i\hbar\partial_t \psi = \mathcal{L}\psi$, about the stationary state, $\sqrt{n}\chi$, using the ansatz 
     \begin{align}
        \psi(\br,t) =&e^{-i\mu t} \Bigl\{\sqrt{n}\chi(\rho) \notag\\
         +\sum_{k_z,m,j}\Bigl[&c_{k_zmj} u_{k_zmj}(\rho)e^{im\phi+i{k_z}z-iE_{k_zmj}t/\hbar} \notag\\
         -&c_{k_zmj}^*v_{k_zmj}^*(\rho)e^{-im\phi-ik_z z+iE^*_{k_zmj}t/\hbar}\Bigr]\Bigr\},
     \end{align}
     where $\phi$ is the azimuthal angle.
Here $\{u_{k_zmj},v_{k_zmj} \}$ are the quasiparticle modes with respective eigenvalues $\{E_{k_zmj}\}$, and $\{c_{k_zmj}\}$ are (small) $c$-number amplitudes.  The $z$-components of linear and angular momentum of the quasiparticle are given by $\hbar k_z$ and $\hbar m$, respectively, and the quantum number $j$ characterizes the transverse (vibrational) mode.     The quasiparticles are obtained by solving the Bogoliubov de-Gennes (BdG) equations  
     \begin{align}\label{bdgm}
      E_{k_zmj}\begin{pmatrix} u_{k_zmj}\\ v_{k_zmj}  \end{pmatrix} = 
      \begin{pmatrix} \mathcal{L}_{k_zm} -\mu& -X_{k_zm}\\ X_{k_zm} & - (\mathcal{L}_{k_zm}-\mu) \end{pmatrix}
      \begin{pmatrix} u_{k_zmj}\\ v_{k_zmj}\end{pmatrix},
      \end{align} 
      where  $\mathcal{L}_{k_zm} \equiv \mathcal{L}^{\perp}  + \epsilon_{k_z}-\frac{\hbar^2m^2}{2M\rho^2} +X_{k_zm}$, $\epsilon_{k_z} =\frac{\hbar^2k_z^2}{2M}$ is the free-particle dispersion relation,   
      and the exchange term is
      \begin{align}\label{x}
      X_{k_zm} f &\equiv n\chi e^{im\phi}\MF^{-1}\{\Ut(\bk_{\rho} + k_z\hat{\mathbf{z}}) \MF\{\chi fe^{-im\phi}\}\}\nonumber \\
      &+ \frac{3}{2}\gamma_\text{QF}n^{3/2}|\chi|^3f,
      \end{align}
      with $\MF$  denoting the planar Fourier transform in $\brho=(x,y)$, and $\bk_\rho=(k_x,k_y,0)$ being the reciprocal space vector.
The excitations explicitly depend on the ${k}$-space interaction potential, and care needs to taken to deal with finite grid size effects in numerical calculations (e.g.~using a cutoff interaction, see Refs.~\cite{Ronen2007a,Lu2010a,Lee2021a}).  
       
\subsection{Variational theory} 
A variational approach can also be used to develop an approximate description of the infinite droplet.  
Here we approximate the transverse mode by a radially symmetric Gaussian  
\begin{eqnarray} \label{transg}
\chi^\text{var}(\rho) = \frac{e^{- \rho^2/2 l^2}}{\sqrt{\pi}l},
\end{eqnarray}
where the width   $l$ is considered as a variational parameter. The energy density per particle of the variational state is (from Ref.~\cite{Pal2020}, but extended to include the quantum fluctuation term)
\begin{eqnarray}\label{Evar}
\mathcal{E} =  \frac{\hbar^2}{2Ml^2} + \frac{M\omega_\rho^2l^2}{2}+\frac{n(g_s-g_{dd})}{4\pi l^2}  + \frac{4n^{3/2}\gamma_\text{QF}}{25\pi^{3/2}l^3}.
\end{eqnarray}  The variational solution is determined by minimising  Eq.~(\ref{Evar}) with respect to $l$.

\begin{figure}[tbh]
	\centering
	\includegraphics[width=3.2in]{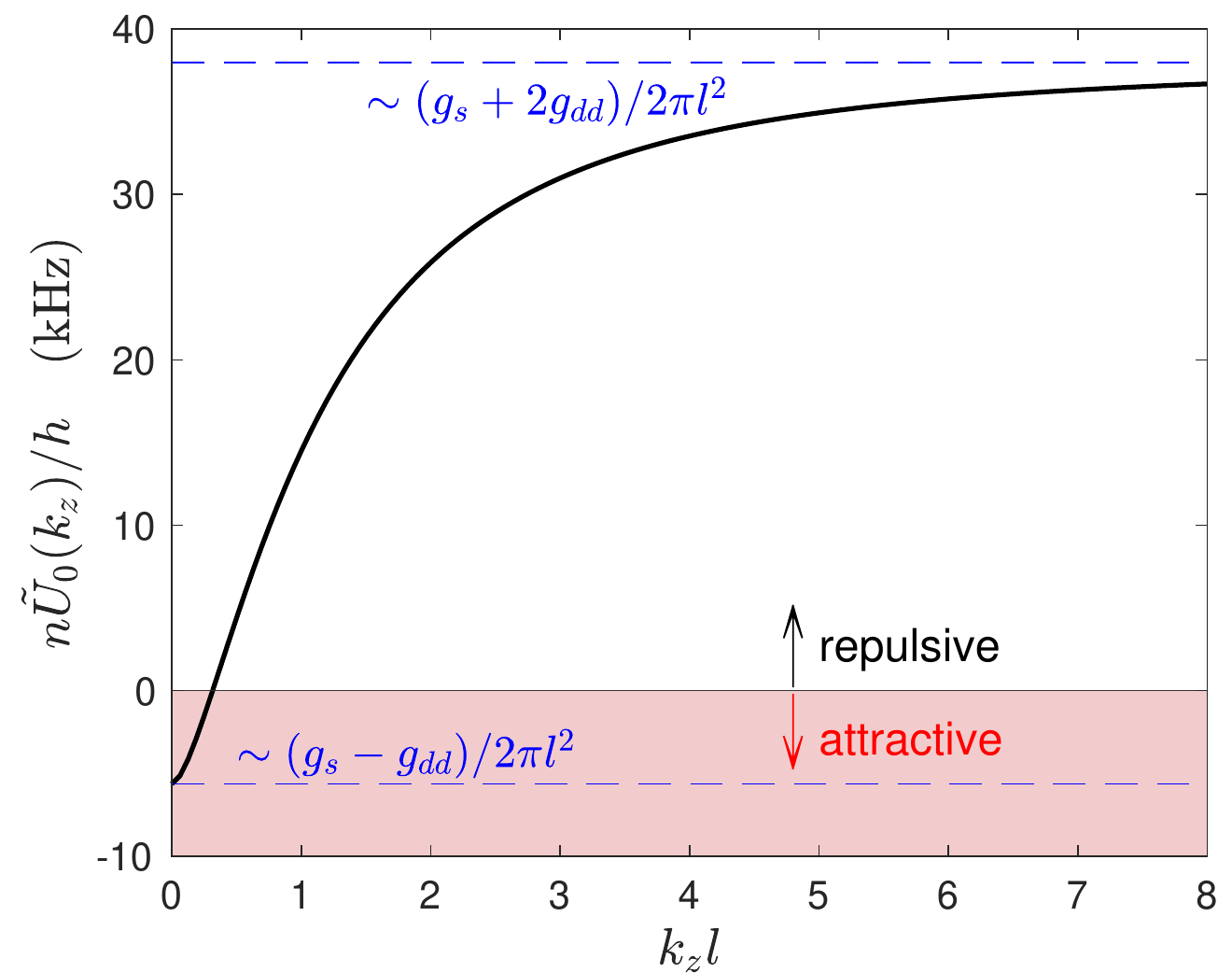}
	\caption{Variational theory effective interaction $\tilde{U}_0(k_z)$ with the low-$k_z$ and high-$k_z$ limits shown. Results for a $^{164}$Dy system with $n=2.5\times10^3/\mu$m,  $a_s=80\,a_0$, $a_{dd}=130.8\,a_0$, and $l=0.383\,\mu$m.}
	\label{Uk0fig}
\end{figure}

The lowest ($m=0,j=0$) band of excitations can be obtained from the variational theory under the same shape approximation  \cite{Pal2020} (also see \cite{Edmonds2020a}), i.e.~setting $u_{k_z00}$ and $v_{k_z00}$  to be proportional to $\chi^\text{var}$. This yields the dispersion relation
\begin{equation} \label{var_E}
E_{k_z}^\text{var}=\sqrt{\epsilon_{k_z}\left[ \epsilon_{k_z}+ 2n\tilde{U}_{0}(k_z) + 3n^{3/2}\gamma_\text{QF} \gamma_\chi\right]},
\end{equation}
where $\gamma_{\chi} \equiv  \frac{2}{5\pi^{3/2}l^3}$, and
\begin{align}
\tilde{U}_0(k_z)=\frac{g_s-g_{dd}}{2\pi l^2} -\frac{3g_{dd}}{2\pi l^2}Q^2e^{Q^2}\text{Ei}(-Q^2),\label{U0kz}
\end{align}  with $Q^2\equiv k_z^2l^2/2$ and $\text{Ei}$ being the exponential integral. 

The behavior of the effective  interaction $\tilde{U}_0(k_z)$ that appears in the variational theory is shown in Fig.~\ref{Uk0fig} for a system in the droplet regime, where $g_{dd}>g_s$. This emphasizes the limiting values of $\tilde{U}_0(k_z)$, notably $\lim_{k_z\to0}\tilde{U}_0=({g_s-g_{dd}})/{2\pi l^2}$ and $\lim_{k_z\to\infty}\tilde{U}_0=({g_s+2g_{dd}})/{2\pi l^2}$. In the droplet regime these limits are attractive and repulsive, respectively.

\section{Results}
\subsection{Dynamic structure factor of an infinite droplet}\label{Sec:DSFinf}

 The zero-temperature dynamic structure factor within Bogoliubov theory is given by
\begin{align}
 S(k_z,\omega)=\sum_j|\delta n_{k_zj}|^2\delta(\omega-E_{k_z0j}/\hbar),\label{DSF}
\end{align}
where  
$\delta n_{k_zj} = \int d{\rho}\,2\pi\rho\big[u_{k_z0j}^{\ast}(\rho) - v_{k_z0j}^{\ast }( \rho)\big]\chi (\rho)$. Here we have considered the dynamic structure factor for a wavevector along $z$ and in this case only the $m=0$ excitations contribute.   
The dynamic structure factor characterizes the structure of the system and its collective excitations. In cold-gas experiments the zero temperature dynamic structure factor can be measured by Bragg spectroscopy (e.g.~see \cite{Stenger1999a,StamperKurn1999a,Blakie2002a}) and this technique has been applied to dipolar condensates in the roton \cite{Petter2019a} and the supersolid \cite{Petter2021a} regimes.

\begin{figure}[tbh]
	\centering
	\includegraphics[width=3.4in]{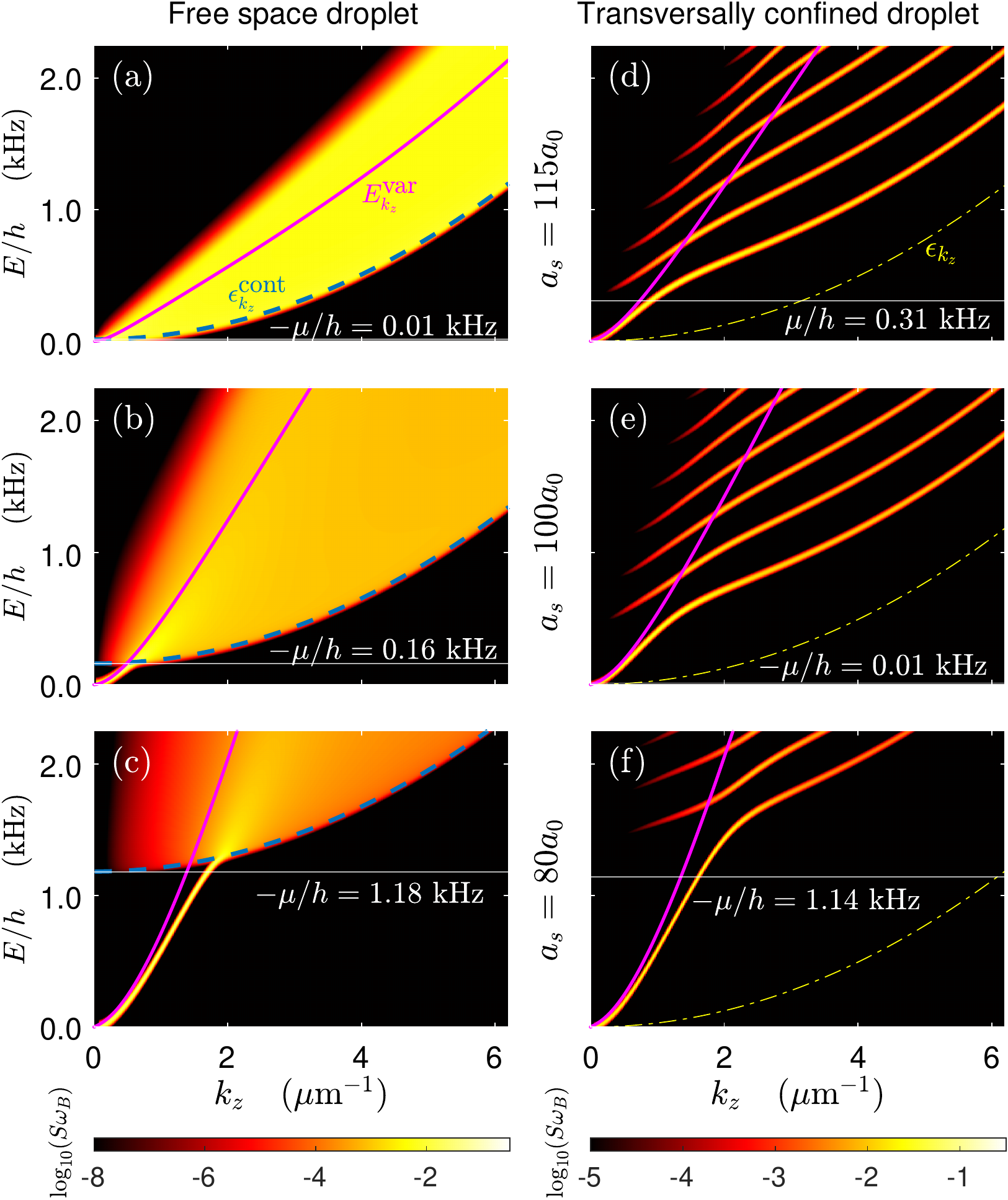}
	\caption{(a)-(c): The dynamic structure factor $S(k_z,\omega)$ for a infinite droplet state of $^{164}$Dy atoms with $n = 2.5\times 10^3/\mu$m and for (a,d) $a_s = 115 a_0$ (b,e) $a_s = 100 a_0$ (c,f) $a_s = 80 a_0$.   Cases (a)-(c) are in free-space ($\omega_\rho=0$) and (d)-(f) are confined with $\omega_\rho/2\pi=150\,$Hz. 
	The variational result  for $E_{k_z}^\text{var}$  is shown with magenta solid lines. 
	 The lower bound of the continuum $\epsilon^\text{cont}_{k_z}$ is shown by the blue dashed lines in (a)-(c), and in (d)-(f) the yellow dash-dot line shows the free-particle dispersion $\epsilon_{k_z}$.
     We broadened the $\delta$ function in the dynamic structure factors by setting $\delta(\omega) = e^{-(\omega/\omega_B)^2}/\sqrt\pi \omega_B$ with $\omega_B/2\pi=15$Hz.
     }
	\label{sb_dsf}
\end{figure} 

In Figs.~\ref{sb_dsf}(a)-(c) we show the dynamic structure factor for free-space droplets with various values of $a_s$. As $a_s$ decreases ($\epsilon_{dd}$ increases) the droplet is more tightly bound, as revealed by the chemical potential becoming more negative. The BdG energies are measured relative to the chemical potential, and the excitations with $E_{k_zmj}<-\mu$ are below threshold and are thus bound to the droplet. Our results show that only one of the $m=0$ branches is bound and it joins up with a continuum of transverse excitations when $E_{k_z00}>\epsilon^\text{cont}_{k_z}\equiv-\mu +\epsilon_{k_z}$. The bound excitation branch is only apparent in Figs.~\ref{sb_dsf}(b) and (c) where $\mu$ is sufficiently negative.

In Figs.~\ref{sb_dsf}(d)-(f) we consider cases with the same parameters used for Figs.~\ref{sb_dsf}(a)-(c), but with the addition  of transverse confinement. For the highest value of scattering length [Fig.~\ref{sb_dsf}(d)] the chemical potential is positive and the system cannot be considered as a self-bound droplet. The observed behavior in this regime is similar to previous work considering a dipolar condensate in an elongated trap (e.g.~see Fig.~3 of  \cite{Pal2020}\footnote{We note that because Ref.~\cite{Pal2020} did not include quantum fluctuations it was restricted to the regime $a_s>a_{dd}$.}). A noticeable effect of the transverse confinement is that discrete excitation bands remain at all energy scales (i.e.~there is no continuum of transverse excitations).  These results also exhibit features which were collectively referred to as the ``anti-roton" effect 
 in Ref.~\cite{Pal2020}. That work considered a dipolar condensate in an elongated trap with the dipoles polarized along the long axis of the trap.
 The features of this effect were identified as:  (i) a rapidly rising, and upward curving, lowest excitation band; (ii) the emergence of a strong multi-band response at high $k_z$. These features arise from the interactions becoming increasingly repulsive\footnote{c.f the roton effect where the interactions are attractive at high $k_z$.} with increasing $k_z$. This situation arises in the droplet where the $k_z\to0$ interaction is attractive allowing the droplet to bind tightly, while as $k_z$ increases the interaction becomes strongly repulsive  [e.g.~see Fig.~\ref{Uk0fig}]. This causes a strong interaction between the respective excitations and the droplet, and is the origin of the multi-band response. 
 In the multi-band regime the transverse profile of the lowest band of excitations changes shape to reduce overlap (and hence interaction) with droplet. For $k_z$ values where this reshaping occurs [e.g.~$k_z\sim1/\mu$m in Figs.~\ref{sb_dsf}(d)-(f)] structure factor response weight  [i.e.~$\delta n_{k_zj}$] is transferred to higher excitation bands.
The variational approach for the lowest excitation band assumes these modes to have the same transverse shape as the droplet, and thus fails to capture the multi-band response. We also observe that aspects of the anti-roton effect carry over to the free-space droplet. First, the ground band curves up rapidly. Second, the multi-band behavior becomes a region of stronger response in the continuum of excitations [see Figs.~\ref{sb_dsf}(b) and (c)].

We note that in all free-space results [Figs.~\ref{sb_dsf}(a)-(c)] and the most strongly bound trap case  [Fig.~\ref{sb_dsf}(f)] that the lowest excitation band $E_{k_z00}$ is imaginary for very small $k_z$ values [e.g.~$\text{Im}\{E_{k_z00}\}$ for the case in Fig.~\ref{sb_dsf}(c) is shown in the lower inset to Fig.~\ref{mc2}(a)]. We neglect the weight of these modes in Fig.~\ref{sb_dsf}, and return to discuss them further in Sec.~\ref{Sec:soundcompressibility}.

\subsection{Excitation spectrum of a finite droplet}\label{Sec:finitedropletexcitations}
  \begin{figure}[tbh]
  	\centering
  	 \includegraphics[width=3.4in]{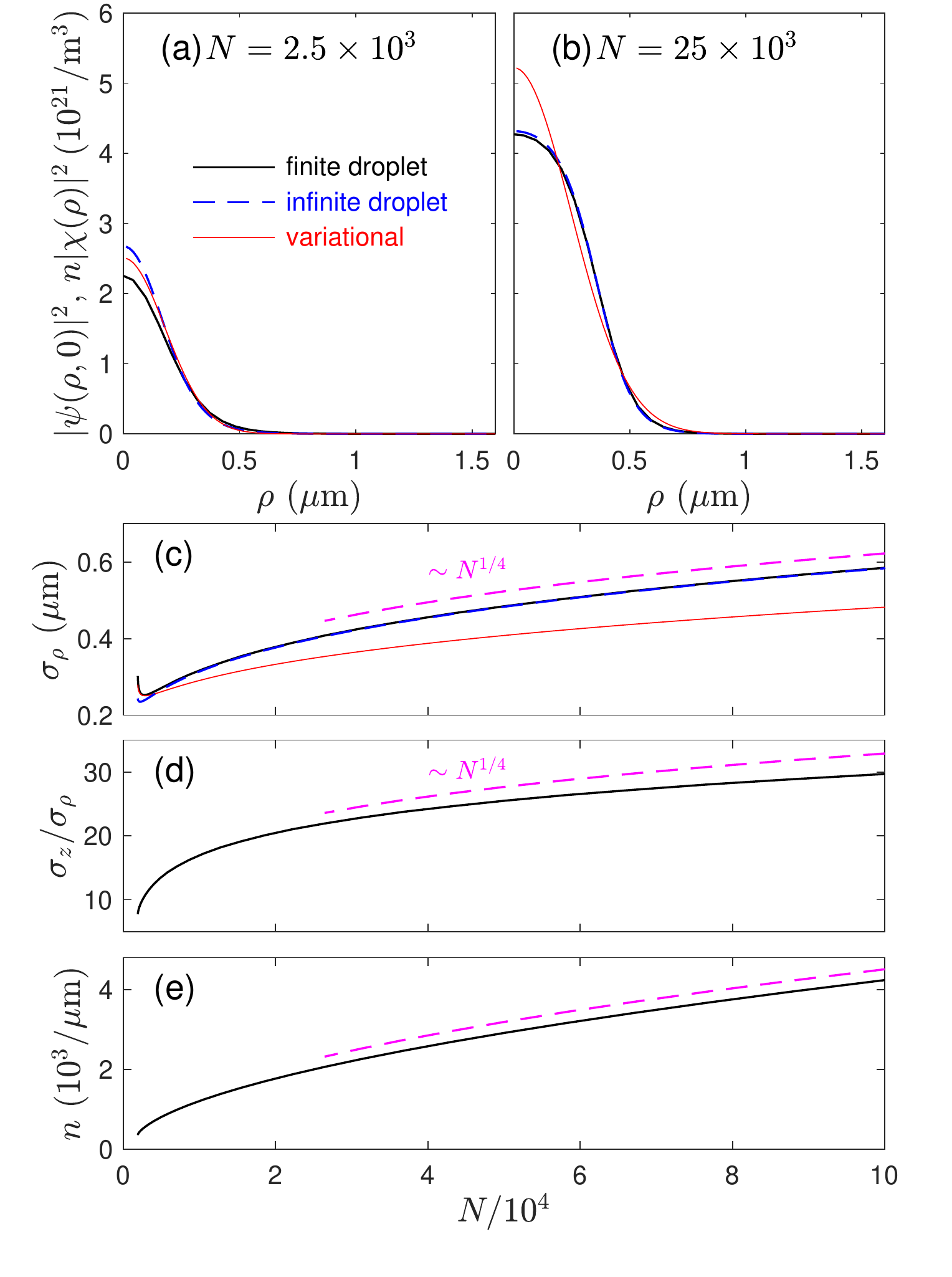}
  	\caption{Mapping and comparison of finite and infinite droplets. The $z=0$ transverse density profile for a finite droplet with (a) $N=2.5\times10^3$ and (b) $N=25\times10^3$ atoms compared to the infinite droplet results. (c) Comparison of the $1/e$-density half width $\sigma_\rho$ in the $z=0$ plane as the number of atoms in the droplet varies (for the variational theory this is $l$). The finite droplet (d) aspect ratio and (e)   linear density (\ref{nmapping}) as $N$ varies.
Here $\sigma_z$ is the $1/e$-density half width of the droplet on the $z$-axis. These results are for a free-space droplet of $^{164}$Dy atoms with $a_{dd}=130.8a_0$ and   $a_s=80a_0$. The droplet is only self-bound for  $N$ exceeding the critical value of $N_c\approx1899$.  }
  	\label{Fig:gsdroplets}
  \end{figure}

We now consider how to apply the infinite droplet theory to characterize the excitation spectrum of a finite droplet.
A finite droplet configuration is specified by the parameters $\{a_s,a_{dd},N\}$.
 Here we take the corresponding infinite droplet to have a linear density identified with the linear density at the centre ($z=0$) of the finite droplet, i.e.~
\begin{align}
n=\int d\brho|\psi(\brho,0)|^2,\label{nmapping}
\end{align}
where $\psi(\brho,z)$ is the wavefunction of the finite droplet.
In Figs.~\ref{Fig:gsdroplets}(a) - (c) we compare aspects of the transverse density profiles obtained using this mapping. Figure \ref{Fig:gsdroplets}(e) shows the linear density as a function of $N$ obtained from finite droplet calculations according to Eq.~(\ref{nmapping}). This linear density is the only additional input into the infinite droplet theory.

These results show that the infinite droplet generally provides a good description of the radial profile except for low atom number cases close to where the droplet unbinds [see Figs.~\ref{Fig:gsdroplets}(a), (c)]. This coincides with the finite droplet having its smallest aspect ratio [Fig.~\ref{Fig:gsdroplets}(d)], where the approximation of the droplet being infinite is least appropriate.  While the infinite droplet theory is seen to work very well for large $N$ (i.e.~the macro-droplet regime),  the variational description works less well in this regime. This is because the density of the droplet saturates towards a maximum value\footnote{Arising from the balance of the attractive two-body interactions and the repulsive quantum fluctuations.}, causing the transverse density profile to develop a flat top that is not well-described by a Gaussian.

\begin{figure}[tbh]
	\centering
	 \includegraphics[width=1.0\linewidth]{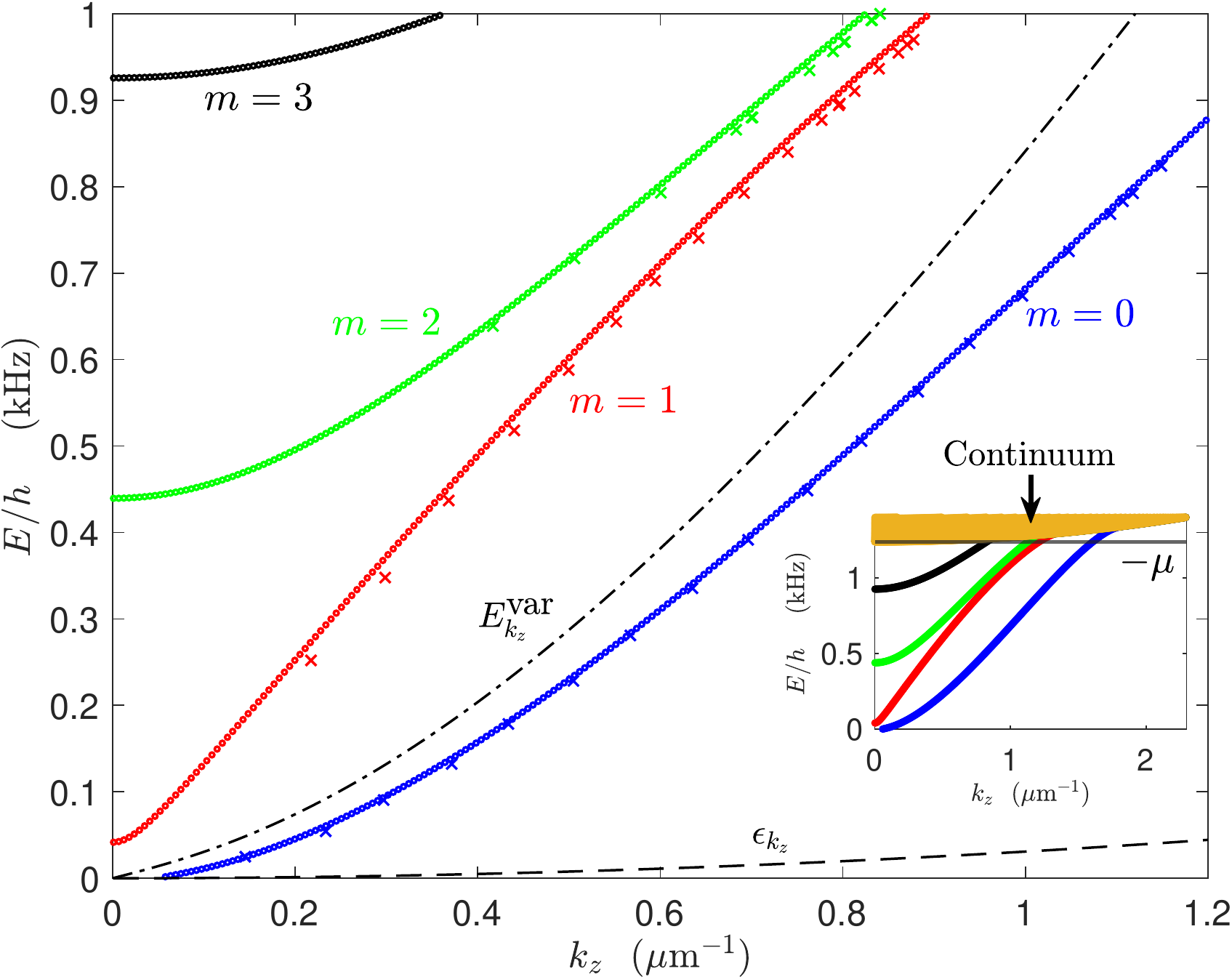}
     \caption{Excitation spectrum of a free-space droplet of $N=10^5$ $^{164}$Dy atoms with $a_s = 80a_0$. The BdG excitation energies of a finite droplet are plotted as $\times$-symbols against a $z$-wavevector assignment (these results are from Ref.~\cite{Baillie2017a} and assignment details are discussed therein). Results are sorted by the $z$-component of angular momentum of the excitation [blue ($m=0$), red ($m=1$), green ($m=2$), black ($m=3$)]. Solid lines give the infinite droplet results for comparison using the linear density $n = 4231.6/\mu$m. Black dash-dotted lines give the variational result Eq.~(\ref{var_E}), and the black dashed line shows  $\epsilon_{k_z}$ for reference. Inset shows the infinite droplet result including the continuum excitations that develop at higher energies. Note that the $E_{k_z00}$ band (blue dots) becomes imaginary for $k_z\lesssim0.04\mu$m$^{-1}$. 
	 }
	\label{fig:finitedropletexcitation}
\end{figure}

 Using our mapping we can provide an approximate description of the excitation spectrum of a finite droplet. Here we compare to results  presented in Ref.~\cite{Baillie2017a} for the spectrum of a finite droplet, obtained using large-scale diagonalization procedures. Some results of that work are reproduced in Fig.~\ref{fig:finitedropletexcitation}. In this regime the droplet is highly elongated ($\sigma_z/\sigma_\rho\approx30$) and acts as a waveguide for the low energy collective modes. 
 The excitations appear as a set of bands labeled by their $z$-component of angular momentum\footnote{In a finite droplet, the $z$-component of momentum is not a good quantum number and in Ref.~\cite{Baillie2017a} an approximate mapping was developed by analysing the approximate wavelengths of the individual excitations.}. In Ref.~\cite{Baillie2017a} an approximate dispersion relation, equivalent to our variational result (\ref{var_E}), was used to provide a qualitative description of the  $E_{k_z00}$ band.  This variational result does not describe any of the higher bands (see Fig.~\ref{fig:finitedropletexcitation}).
 In contrast the infinite droplet theory [numerically solving Eqs.~(\ref{bdgm})] exhibits good quantitative agreement with the finite droplet results for the $m=0,1,2$ bands of excitations presented in Ref.~\cite{Baillie2017a}. We also show the infinite droplet predictions for the $m=3$ band (not calculated in Ref.~\cite{Baillie2017a}), with the higher angular momentum bands lying in the continuum (see the inset to Fig.~\ref{fig:finitedropletexcitation}). We note that the finite droplet theory predicts that the $E_{k_z00}$ band is imaginary in the small $k_z$ limit (well below where the first excitation occurs in the finite droplet).

\subsection{Axial speed of sound and long-wavelength instability of the infinite droplet}\label{Sec:soundcompressibility}
The excitation spectrum of a dipolar condensate is known to be anisotropic (e.g.~see \cite{Wilson2010a,Ticknor2011a,Bismut2012a,Baillie2014a,Wenzel2018a}) due to the anisotropy of the DDIs. Here we explore the axial speed of sound of a dipolar droplet - i.e.,~the speed of sound along the long axis of the droplet. 
We identify the axial speed of sound from the long-wavelength slope of the lowest excitation band as
\begin{align}
c_z=\frac{1}{\hbar}\left.\frac{\partial E_{k_z00}}{\partial k_z}\right|_{k_z=0}.\label{sosBdG}
\end{align} 
The $E_{k_z00}$ excitation band was observed in Figs.~\ref{sb_dsf} and \ref{fig:finitedropletexcitation} to have novel behavior arising from the anti-roton effect, notably a rapid rising and upwardly curving shape. These features make identifying the speed of sound difficult from a visual inspection of the results. Furthermore, in the previous subsections we noted that at the lowest band can be dynamically unstable at very low $k_z$, which corresponds to $c_z$ being imaginary (i.e.~$Mc_z^2<0$).

 The variational spectrum in Eq.~(\ref{var_E}) neglects any change in the transverse profile for long wavelength excitations and thus does not provide a good estimate of the speed of sound. So for the variational approach it is better to use the result  $Mc_z^2=n(\partial n/\partial \mu)$ for the inverse (axial) compressibility\footnote{Note we can also use this result to obtain $Mc_z^2$ directly from the EGPE rather than calculating the BdG excitations. We have verified that the results are identical from these two approaches.}
 \begin{align} \label{mc2var}
M&c^2_\text{var} =n\left(\frac{\partial ^2\mathcal{E}}{\partial n^2}\right),\\ &=  n\tilde{U}_0(0) \left(1-\frac{n}{l}\frac{dl}{dn} \right) + \frac{3n^{3/2}\gamma_\text{QF}\gamma_\chi}{2}\left(1-\frac65\frac{n}l\frac{dl}{dn}\right),
\end{align} 
where $\tilde{U}_0(0)={(g_s-g_{dd})}/{2\pi l^2}$ [from Eq.~(\ref{U0kz})],  and
\begin{align}
\frac{n}l\frac{dl}{dn}=\frac{   n\tilde{U}_0(0)+\frac95n^{3/2}\gamma_\text{QF}\gamma_\chi }{  \frac{3\hbar^2}{ml^2}+\frac{ml^2\omega_\rho^2}{2}+ 3n\tilde{U}_0(0) +\frac{24}5n^{3/2}\gamma_\text{QF}\gamma_\chi }.
\end{align}

       \begin{figure}[tbh]
       	\centering
       	 \includegraphics[width=1.0\linewidth]{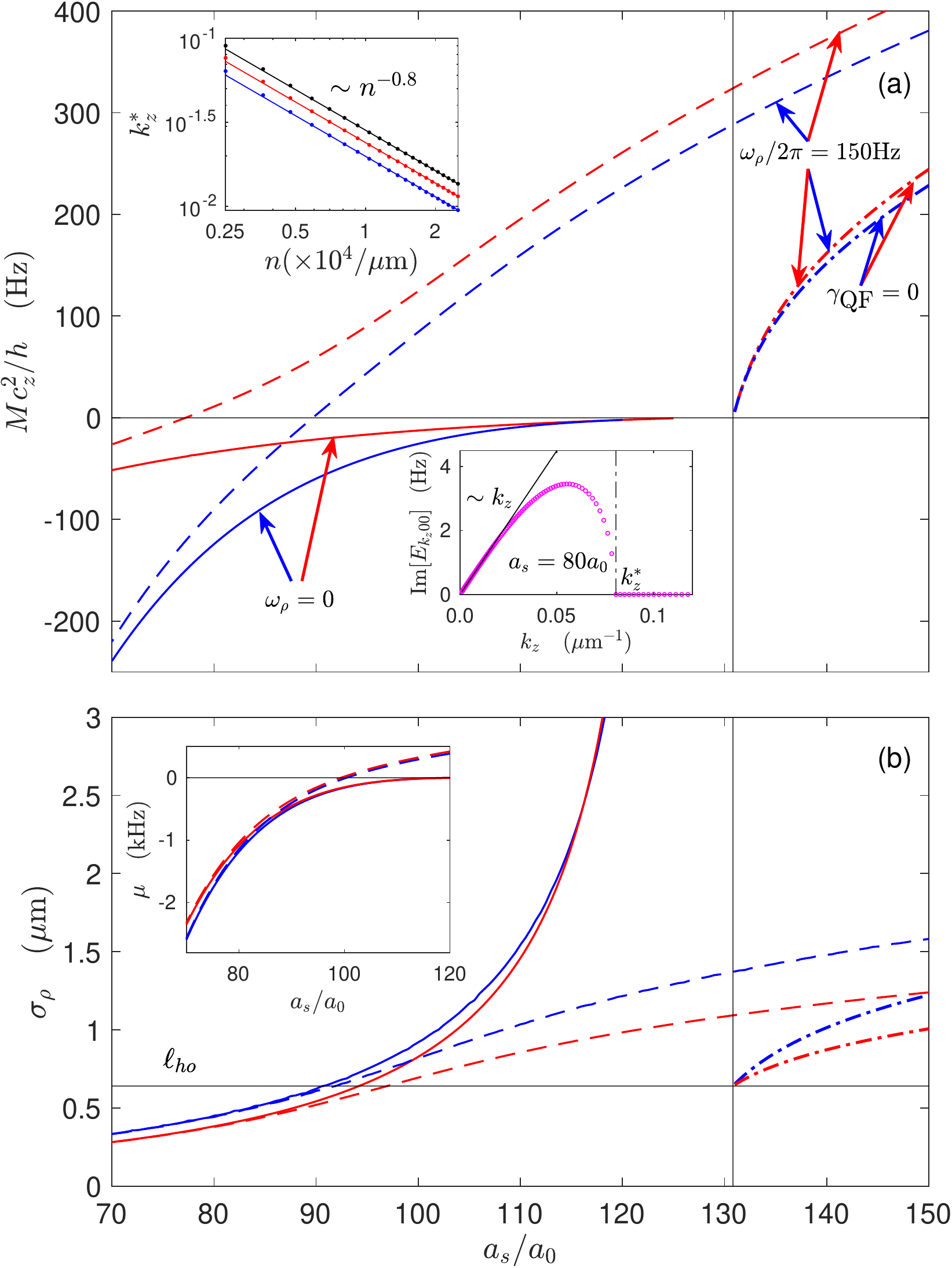}
       	\caption{ (a) $Mc_z^2$ \cite{Pal2020} as a function of $a_s$ for fixed linear density $n = 2.5\times 10^3/\mu$m from EGPE (blue lines) and variation (red line) calculations. Results are shown for (solid) infinite free-space droplets; (dashed) transversally confined infinite droplets; and (dashed dot) a transversally confined condensate with $\gamma_\text{QF}=0$.  The upper inset shows how $k_z^*$ varies with $n$ for $a_s/a_0 = 75$ (black, top), $80$ (red), $85$ (blue). The lower inset shows the imaginary part of the $E_{k_z00}$ spectrum for a free-space droplet with $a_s = 80a_0$.   (b) The radial $1/e$-density widths of the states.  These results  use $a_{dd}=130.8a_0$, and for trapped cases $\omega_\rho/2\pi=150\,$Hz.
	}
       	\label{mc2}
       \end{figure} 
       \vspace{0.5cm}

In Fig.~\ref{mc2}(a) we plot $Mc_z^2$ as a function of $a_s$ for infinite droplets in free-space  and with transverse confinement.  For reference, we also show the behavior of a transversally confined system without quantum fluctuations, where  $Mc_z^2\to0$ (and the system collapses) as $a_s\to a_{dd}$ from above.  Comparing the trapped results we observe that inclusion of quantum fluctuations causes a significant increase in $Mc^2_z$. However, even including quantum fluctuation effects $Mc^2_z$ becomes negative for sufficiently low values of $a_s$ (i.e.~$a_s\lesssim90a_0$). For the infinite free-space droplet at all values of $a_s$ where a localized droplet state is obtained we find that $Mc^2_z<0$.

A negative value of $Mc^2_z$, i.e.~when $c_z$ is imaginary, indicates that the long wavelength modes of the system are dynamically unstable. The lower inset to Fig.~\ref{mc2}(a) shows the imaginary part of the spectrum for an infinite free-space droplet with $a_s=80a_0$. Here the $E_{k_z00}$ excitation band is purely imaginary for $k_z<k_z^*$ and real thereafter.  
For the finite sized droplet the longest wavelength excitation has a wave vector of approximately $k_{z,\min}\equiv\pi/\sigma_z$ (cf.~the first $m=0$ excitation marked by a symbol in Fig.~\ref{fig:finitedropletexcitation}), where $\sigma_z$ is the $1/e$ density half-width introduced earlier. Thus the finite axial extent of the droplet prevents the unstable modes being accessed, so the dynamical instability cannot manifest.  The upper inset to Fig.~\ref{mc2}(a) shows results for how the maximum unstable wave vector, $k_z^*$, changes with $a_s$ and $n$. For larger finite droplets, where $n$ is higher,  $k_z^*$ gets smaller, possibly indicating that the instability cannot manifest even in very large droplets. This is consistent with the observations in Ref.~\cite{Baillie2017a} where the lowest energy $m=0$ mode of a finite droplet appears to asymptotically approach zero as $N$ increases.

The width $\sigma_\rho$ of the infinite free-space droplet diverges with increasing $a_s$ [see Fig.~\ref{mc2}(b)], as the droplet unbinds and evaporates.
Using the variational theory we can solve for the evaporation transition, i.e.~where a stationary state solution occurs with $\mathcal{E}=0$, giving
\begin{align}
\frac{1}{2} + n(a_s-a_{dd}) = 0. \label{EvapCond}
\end{align} 
For the free-space droplet case in  Fig.~\ref{mc2} this result predicts evaporation to occur at $a_s=127a_0$. Because the radial width diverges as we approach the transition we have not been able to numerically explore the infinite droplet EGPE predictions for the transition in detail.  
 For a finite droplet as $\sigma_\rho$ increases the infinite droplet becomes a poor approximation, since it requires $\sigma_z\gg\sigma_\rho$. Thus, condition (\ref{EvapCond}) is not a useful predictor of the transition in a finite droplet.

\subsection{Thermodynamic limit droplet of a free-space droplet}

Finally we consider the thermodynamic limit of a free-space dipolar droplet. We take the thermodynamic limit  by allowing $N$ to increase towards infinity in a finite free-space droplet.  We find (e.g.~from fits in Fig.~\ref{Fig:gsdroplets}(c)-(e), also see Ref.~\cite{Baillie2017a}) that the following approximately scaling behavior  holds
\begin{align}
\sigma_\rho&\sim N^{1/4},\\
\sigma_z&\sim N^{1/2},\\
n&\sim\sigma_\rho^2\sim N^{1/2},
\end{align}
for the width, length and linear density of the droplet, respectively.  Thus, in the thermodynamic limit all of these quantities are infinite. Hence, the thermodynamic limit does not correspond to the infinite droplet theory developed in this paper, in which $\sigma_z$ is infinite, but $\sigma_\rho$ and $n$ are both finite.

\section{Conclusions and outlook}
       \label{sec:conclusions}

 In this paper we have developed a theory to describe an infinitely long dipolar droplet, as an idealization of the long filament shaped droplets prepared in experiments. Our focus has been on the excitation properties, which are difficult to calculate accurately for finite droplets. 
  Our results in Fig.~\ref{fig:finitedropletexcitation}  demonstrate good quantitative agreement to one of the few detailed studies of a finite dipolar droplet excitations in the literature. 
In the excitation spectrum, a number of interesting features arise, collectively called the anti-roton effect \cite{Pal2020}.   
These features appear in our results for the dynamic structure factor of an infinite droplet, which could be measured in experiments by performing Bragg spectroscopy along the long axis of a large droplet.

 We have also examined the nature of the axial speed of sound in a droplet. For the infinite droplet theory we find that $Mc^2_z<0$ for free-space droplets and in confined droplets with sufficiently low values of $a_s$. 
 The $k_z\to0$ limit used to identify the speed of sound is problematic for finite droplets due to prominent finite size dependence arising from the anti-roton effect (i.e.~the rapid change in the effective interaction with $k_z$ near $k_z=0$). Indeed, finite dipolar droplets appear to be stable because the unstable modes occur at wavelengths that are too long to be accommodated. Increasing $N$ to make the droplet longer, also causes $n$ to increase (the droplet gets wider), and our results show that this leads to unstable wavelengths becoming even longer. Thus it is possible that these instabilities never manifest in finite droplets, and certainly this is consistent with calculations we have performed for finite droplets over a wide parameter regime. This also raises the question: what is the appropriate axial speed of sound for a finite droplet? In practice a result reflecting the finite size of the system may be necessary, e.g.~replacing $k_z=0$ with $k_z\to1/\sigma_z$ in Eq.~(\ref{sosBdG}). Whether such an identification is useful in understanding the effective compressibility or critical velocity of a finite droplet is an interesting direction for future research.

\section{Acknowledgements}
We acknowledge support from the Marsden Fund of the Royal Society of New Zealand, and valuable discussions with the Ferlaino group in Innsbruck and L.~Chomaz.
     
% \bibliography{dipolarpbb} 


      \end{document}